\documentclass[aps,prl,twocolumn,groupedaddress]{revtex4-2}
\usepackage{chemarr}
\usepackage{mathtools}
\usepackage{graphicx,color}
\usepackage[export]{adjustbox}
\usepackage[strings]{underscore}
\usepackage[caption=false]{subfig}
\usepackage{float}
\usepackage{cleveref}
\usepackage{underscore}
\usepackage{xcolor}

\begin{document}
\title{\textcolor{black}{Arrested Ostwald Ripening in Non-Equilibrium Systems}}
\author{Bradley D. Keister}
\affiliation{Center for Astrophysics and Space Sciences, UCSD, La Jolla, CA 92093, USA}
\author{Saeed Najafi}
\affiliation{Department of Chemistry, UCSB, Santa Barbara, CA 93106, USA}
\author{Joan-Emma Shea}
\affiliation{Department of Chemistry, UCSB, Santa Barbara, CA 93106, USA}
\author{Krastan B. Blagoev}
\affiliation{Physics Division, National Science Foundation, Alexandria, VA 22314}
\affiliation{Department of Biophysics, Johns Hopkins University, Baltimore, MD 21218, USA}
\affiliation{Institute of Molecular Biology, Bulgarian Academy of Sciences, Acad.\ G.\ Bonchev Str.\ Bl.\ 21, 1113 Sofia, Bulgaria}
\affiliation{Institut Curie, PSL Research University, Sorbonne Université, CNRS UMR3664, Laboratoire Dynamique du Noyau, 75005 Paris, France}

\date{\today}

\begin{abstract}
Absence of Ostwald ripening has been observed in living cells, which operate out of equilibrium. Using molecular dynamics we study the behaviour of liquid droplets away from equilibrium in a system of particles interacting with each other via several central potential. The system is driven out of equilibrium either by the particles randomly transitioning between two states, or by randomly changing their momenta. In all cases Ostwald ripening is absent only away from equilibrium. This phenomena, might be the mechanism by which droplets in living cells are stabilized.
\end{abstract}

\maketitle

\section{introduction}
Droplet-like organelles  have been observed as distinct membrane-less condensates composed of proteins and RNAs in living cells; examples include stress granules, Cajal bodies, and P-bodies~\cite{Brangwynne2009,Shin2017,VanTreeck2019,Carrettiero2022,Sawyer2016}. It has been proposed that these condensates form via liquid–liquid phase separation (LLPS) and display hallmark features of condensed matter droplets, including fusion, surface tension, and internal molecular rearrangement~\cite{Brangwynne2011,Hyman2014}. However, unlike equilibrium solutions, many biological condensates are maintained in nonequilibrium states by continuous energy input from processes like ATP hydrolysis, active transport, and enzymatic activity~\cite{Patel2017,Alberti2019}. These active processes lead to complex behaviors such as arrested coarsening, asymmetric partitioning, and compositionally driven phase switching\cite{Zwicker:2014a}.

After nucleation, droplets below a critical size (which depends on thermodynamic conditions) disappear, while droplets above that size grow. This behavior is a consequence of the competition between the surface and volume contributions to the Gibbs free energy. The surface of a droplet increases the energy, while the bulk of the droplet lowers the potential energy. The Gibbs–Thomson effect—where the chemical potential increases with droplet curvature—creates a chemical potential gradient between small and large droplets. This gradient drives diffusive particle transport from smaller to larger droplets, leading to coarsening of the supersaturated liquid~\cite{IlyaMLifshitz1961a, Alikakos2004}. This process is further enhanced by droplet coalescence~\cite{Xu2023,Jin2017}. In a liquid dissolved in another liquid, nucleation occurs at lower concentrations when impurities are present, while in a pure liquid–liquid solution, large density fluctuations drive nucleation~\cite{Hyman2014a}.

Droplet formation and dynamics in in vivo systems (and even in some in vitro) do not necessarily obey the principles of equilibrium thermodynamics~\cite{Kolimi2021,Miangolarra2021}. Such out-of-equilibrium behavior can arise from transient or asymmetric interactions that droplet constituents experience either inherently or in response to external stimuli~\cite{Amano2022,Belan2021,Ornes2017,Fodor2022}. In living cells, droplets often do not exhibit Ostwald ripening~\cite{Nakashima2021}, at least not on physiological time scales, and are capable of disassembling when no longer needed. These observations suggest that non-equilibrium conditions may not only arrest phase separation but also stabilize finite-size condensates~\cite{Berthin2024a} and prevent the formation of large bulk phases typical of equilibrium systems.

For example, it has been shown that shear flows and active interfacial stresses can destabilize or even fragment the large droplets that would normally arise in equilibrium~\cite{Weber2019}. Long-range ~\cite{Muratov2002a}  as well as chemical reactions can also lead to steady-state pattern formation ~\cite{Glotzer1994a,Christensen1996a}. Moreover, the presence of chemically active centers in solution can result in droplets with abnormal solubility~\cite{Zwicker2015sor,Zwicker:2014} and be used for droplet size control~\cite{Kirschbaum2021a} . In the context of active colloidal fluids, several studies have investigated droplet formation mechanisms under non-equilibrium conditions~\cite{PhysRevX.8.031080,PhysRevLett.110.055701,PhysRevLett.110.238301}. Recently it was shown that that sticker-spacer condensates can arrest droplet merging at low densities~\cite{Shakhnovich2025a}.

Previous work has shown that active interfacial stresses—and more broadly, processes driving the system out-of-equilibrium—can arrest phase separation~\cite{Tayar2021,Singh2019,PhysRevX.8.031080,Cates2010,Cates2003}. Active forces may also suppress phase separation via self-stirring mechanisms that destabilize the formation of large droplets~\cite{Caballero2022}. Recently, a molecular dynamics study of a model of interacting particles showed that non-equilibrium dynamics can yield small, stable droplets. This model employed a three-body potential in which particle interactions switch from an attractive Lennard-Jones interaction potential to a short-range repulsive force for short-range pair distance~\cite{Rabin:2019}.

In this paper, we explore several forms of two-body interaction potentials and demonstrate that Ostwald ripening is arrested in all cases when the system is driven out of equilibrium. 
Our results suggest that the arrest of Ostwald ripening may be a robust feature of driven nonequilibrium systems, largely independent of the specific form of the interparticle interactions~\cite{Zwicker:2014a,Wurtz2018,Weber2019}.

\section{the models}
A general theory describing condensates out of equilibrium is currently lacking. To explore the underlying physical mechanisms leading to condensed states of matter, molecular dynamics simulations employing Lennard-Jones-type potentials are useful~\cite{Rabin:2019}. Lennard-Jones systems out of equilibrium, can bring insights into the formation, kinetics, and stability, of biomolecular condensates under physiological stress. To advance understanding in this area, we employ molecular dynamics simulations using Lennard-Jones type potentials to study a minimal model of a particle system driven out of equilibrium. In this model, particles can exist in one of two internal states (A or B), and we consider two mechanisms that transfer energy in or out of the system, thereby driving it out of equilibrium.

To drive the system out of equilibrium we consider an abstract cellular environment, where energy input from ATP can promote transitions from \( A \to B \), resulting in the formation of ADP, while the reverse transition \( B \to A \) leads to the regeneration of ATP.

In the first case, the system is driven out of equilibrium by randomly switching the interaction potential between particles in states A and B. In the second case, the system is driven by randomly increasing or decreasing the velocity of a fraction of particles in state B, within the total population (A + B).

Transitions between states occur with a finite probability and drive the system out of equilibrium. When transitions are prohibited, the system remains in equilibrium and, following nucleation, coarsens via Ostwald ripening or droplet coalescence. In contrast, when transitions are allowed, the system evolves to a non-equilibrium steady state characterized by a greater number of smaller droplets. In this state, Ostwald ripening is suppressed, and coalescence becomes the only—though infrequent—mechanism of droplet growth.



%

The particles interact within a three-dimensional isolated box; that is, no heat or particles are exchanged with the external environment. The relative fraction of particles in states $A$ and $B$ is held fixed. The $A-A$ and $A-B$ interactions are modeled using the standard Lennard-Jones potential:

\begin{equation} \label{eq:15} V_{AA}(r) = V_{AB}(r) = 4\epsilon \left[ \left(\frac{\sigma}{r}\right)^{12} - \left(\frac{\sigma}{r}\right)^{6} \right], \end{equation}while the $B-B$ particles interactions are via either a modified Lennard-Jones potential or a double-well potential. In the context of protein folding, a double-well potential has been used to represent interactions between molecules separated by water~\cite{Cheung2002a}.  Here $r$ is the interparticle distance,
$\sigma$ is a characteristic length, and $\epsilon$  is an energy scale which are defined below.
Here, the double-well potential is modeled as a Lennard-Jones potential plus a shifted Gaussian:

\begin{equation} \label{eq:16} V_{BB}(r) = V_{\text{LJ}}(r) + \frac{H}{\delta\sqrt{2\pi}} \exp\left[-\frac{(r - a)^2}{2\delta^2}\right]. \end{equation}

The use of a double-well potential is also motivated by studies of the two-liquid model of supercooled water~\cite{Mishima1998} and its application to polyamorphous substances~\cite{Franzese2001}.

The parameters used in terms of Lennard-Jones reduced units are
where $\epsilon=\sigma=1$, and
\begin{equation}
  \label{eq:3}
  \begin{aligned}
    T^* &= kT / \epsilon; \\
    \rho^* &= \rho \sigma^3.
  \end{aligned}
\end{equation}

\color{black}
\noindent For the shifted Gaussian, we used $H=-0.7$ for the strength of the Gaussian  and 
$\delta=0.2$ for the width of the Gaussian , and values of 1.5, 2.2, and 3.5 for the center of the maximum of the Gaussian $a$.  We these
parameters, there is a competition between the minimum of the Lennard-Jones
potential and that of the Gaussian.  Otherwise one of the two wells
dominates, and the special results that we discuss below do not occur.
\color{black}

We also studied a single-well potential for the $B-B$ interaction that is five
times the $A-A$ Lennard-Jones potential
\begin{equation}
  \label{eq:2}
 V_{BB}(r) = 5 V_{AA}(r).
\end{equation}

These potentials all feature a $B-B$ potential that has more
attraction than the basic Lennard-Jones potential, either from
longer range, or deeper well.

\begin{figure}[h]
  \centering
  \includegraphics[width=1.0\linewidth]{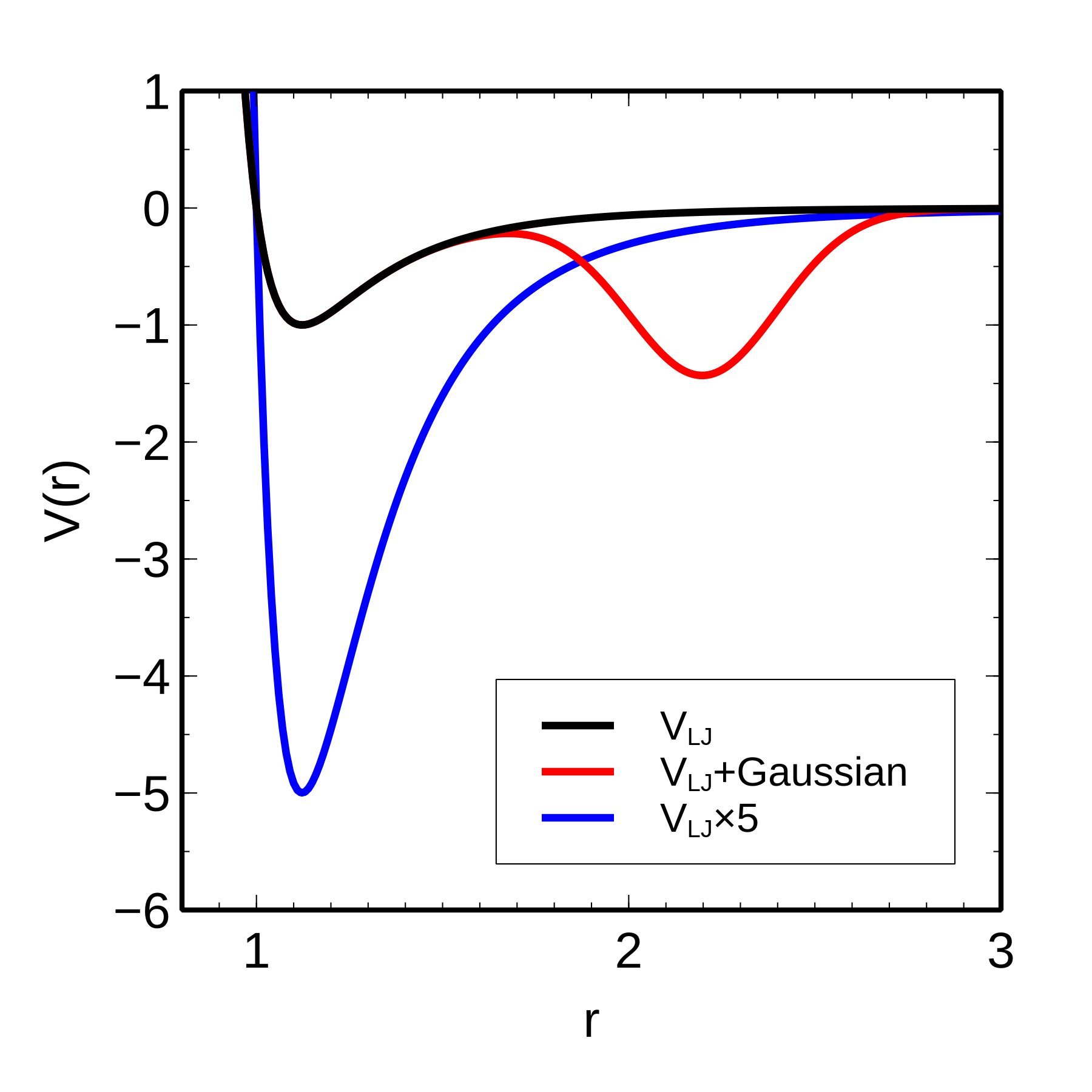}
  \caption{Potentials used in the simulation.  Black: pure Lennard-Jones; red:
    Lennard-Jones+Gaussian; blue: Lennard-Jones times 5.}
  \label{fig:triplet}
\end{figure}

The potentials discussed above are shown in Fig.~\ref{fig:triplet} 

At the density and temperature considered here, the system is weakly
coupled, because the ratio of the average potential energy to the
average kinetic energy is much less than one i.e.  $\Gamma=U/kT\ll 1$,
where $U$ is the average potential energy calculated for each
inter-particle potential at the average inter-particle distance
$r_0=(\frac{3}{4\pi \rho})^{1/3}$ at the studied density $\rho$.

At the density and temperature considered here, the system is weakly coupled, as the ratio of the average potential energy to the average kinetic energy is much less than one; that is,
\begin{equation}
    \Gamma = \frac{U}{kT} \ll 1,
\end{equation}
where \( U \) is the average potential energy, calculated for each inter-particle potential at the average inter-particle distance, given by
\begin{equation}
    r_0 = \left( \frac{3}{4\pi \rho} \right)^{1/3},
\end{equation}
with \( \rho \) the particle number density.

A motivation for this study is the behavior of biological condensates observed in living cells. A ``spherical protein'' composed of, for example, 526 amino acids can be modeled as a hard sphere with a radius (in angstroms) given by~\cite{Dill17876}
\begin{equation}
   R = N^{2/5} = 12.25~\mathrm{\AA},
\end{equation}
which yields a Lennard-Jones spatial parameter
\begin{equation}
    \sigma = 2R = 2.45~\mathrm{nm}.
\end{equation}
Assuming a particle mass \( M = 1.2 \times 10^{-22}~\mathrm{kg} \) and concentration \( \rho = 2.4~\mathrm{nm}^{-3} \), the characteristic time scale of the simulations is
\begin{equation}
    \tau = 4.1 \times 10^{-10}~\mathrm{s},
\end{equation}
implying that the full simulation corresponds to a physical time interval of approximately 1 microsecond.
We studied different volume fractions of the two states and here typical results are shown for 70\% of the particles in state \( A \) and 30\% in state \( B \).
We found that the system must be in a dilute state to avoid very rapid ripening. The results reported here use a reduced temperature \( T^* = 1.25 \) and reduced concentration \( \rho^* = 0.05 \). Here the reduced parameters are defined in terms of the Lennard-Jones units in Equation \ref{eq:3}.

\begin{figure}[h]
\centering
\includegraphics[width=3.5in]{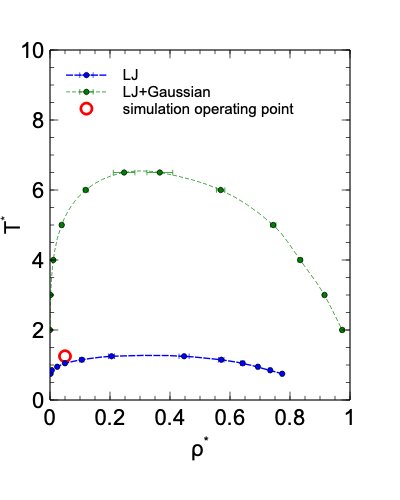}
\caption{Equilibrium phase diagram obtained using Monte-Carlo
  simulations for the Lennard-Jones potential and a pure double-well potential with $a=2.2$.}
\label{fig: MC}
\end{figure}

The concentration and temperature for the MD simulations was chosen to lie outside the liquid--gas coexistence region of an equilibrium Lennard-Jones system.
To verify that the hybrid \( A\!-\!B \) system resides in the condensed phase at equilibrium, we performed Monte Carlo simulations~\cite{DLMONTE} of the system at equilibrium, as shown in Fig.~\ref{fig: MC}, for both the standard Lennard-Jones potential and the double-wel potential with \( a = 2.2 \). 
The MD simulations were performed at a point in the concentration-temperature phase diagram that lies outside the liquid-gas coexistence region for the plain Lennard-Jones system (composed solely of particle A), but inside the coexistence region for particles of type B.  These two borders are shown in Fig.~\ref{fig: MC}, along with the chosen point. 

The operating point for the simulations is indicated in red. Under the chosen conditions, droplets form as collections of \( B \) particles.

 In our model, the system is driven out of equilibrium by forcing particles to switch between the two states with rates \( k_1 \) and \( k_2 \):
\begin{equation}
  \label{eq:22}
  A \xrightleftharpoons[k_2]{k_1} B.
\end{equation}

The corresponding kinetic equations governing the state concentrations are
\begin{equation}
  \label{eq:1}
  \frac{d[B]}{dt} = -\frac{d[A]}{dt} = k_1 [A] - k_2 [B],
\end{equation}
where square brackets denote concentrations.

An equivalent way to implement these transitions in the simulation is to randomly exchange the identities of selected pairs of \( A \) and \( B \) particles at regular time intervals.

When a particle changes its interaction potential with its neighbors, it experiences a different potential energy compared to the value before the transition. This violates local energy conservation and keeps the system out of local equilibrium. However, if the \( A \leftrightarrow B \) transitions occur at equal rates, the system reaches a non-equilibrium steady state.

The diffusive molecular dynamics simulations were performed using the LAMMPS molecular dynamics package~\cite{LAMMPS} in a cubic box containing 27,000 particles. The system evolves under constant number, volume, and energy (NVE) conditions, with a Langevin thermostat applied to bring the system to a steady state. Here the thermostat regulates temperature but does not introduce significant heat exchange beyond what is inherent to the system's non-equilibrium nature. The only energy exchange occurs during the particle transitions, which on average is zero.  Simulations were run for one million time steps with a reduced time step size \( t^* = 0.01 \).

We verified that an equilibrium system of pure \( A \) particles at the chosen concentration does not form droplets. For the non-equilibrium system, 10\% (810) of the \( B \) particles were randomly exchanged with \( A \) particles every 1,000 steps. This exchange protocol ensures a complete turnover of \( B \) particles multiple times over the course of the simulation.

\begin{figure}[h]
\centering
\includegraphics[width=1.0\linewidth]{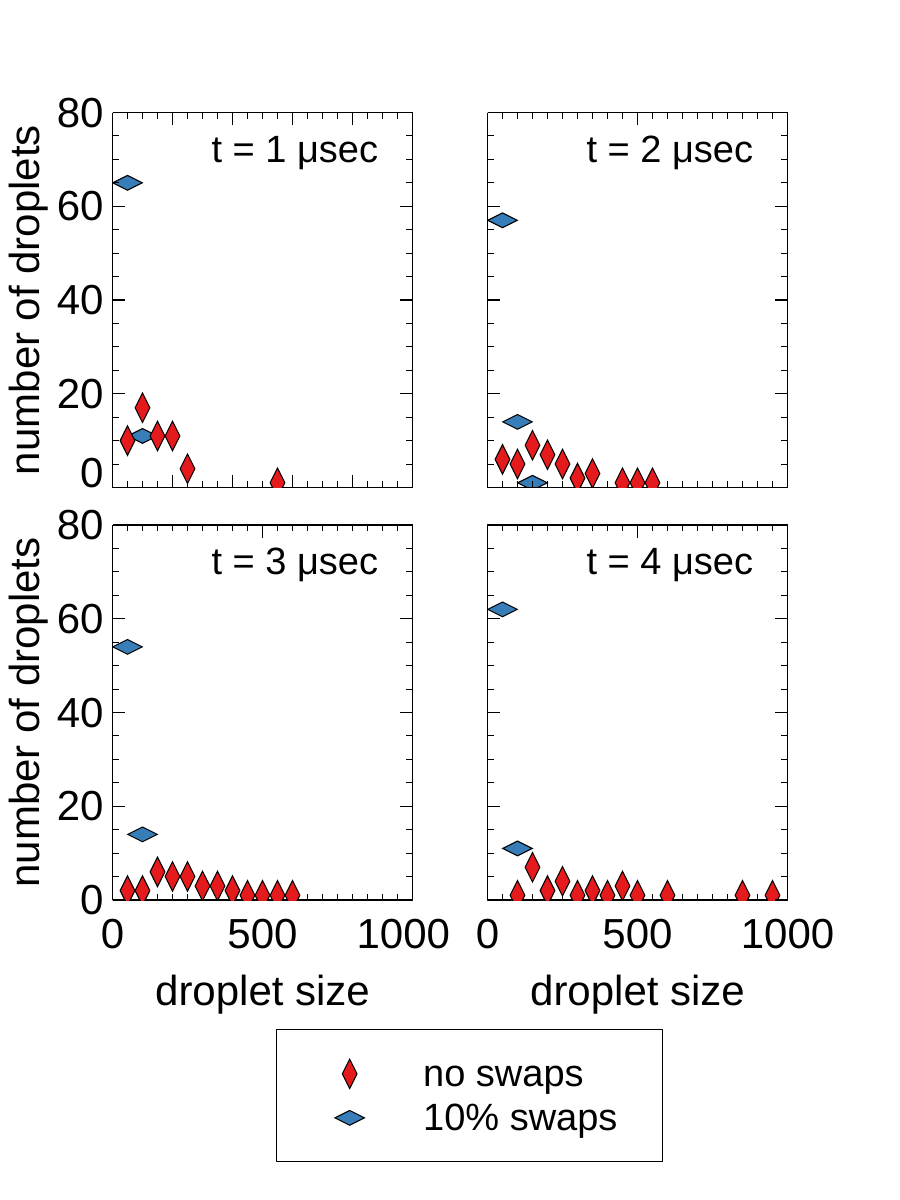}
\\[\smallskipamount] 
\caption{The four panels show the  droplet size distribution at
  various stages of the simulation, with equilibrium results shown in
  red, and non-equlibrium results shown in blue.}
\label{fig:LJDW2p2}
\end{figure}

\begin{figure}[h]
\centering
\includegraphics[width=1.0\linewidth]{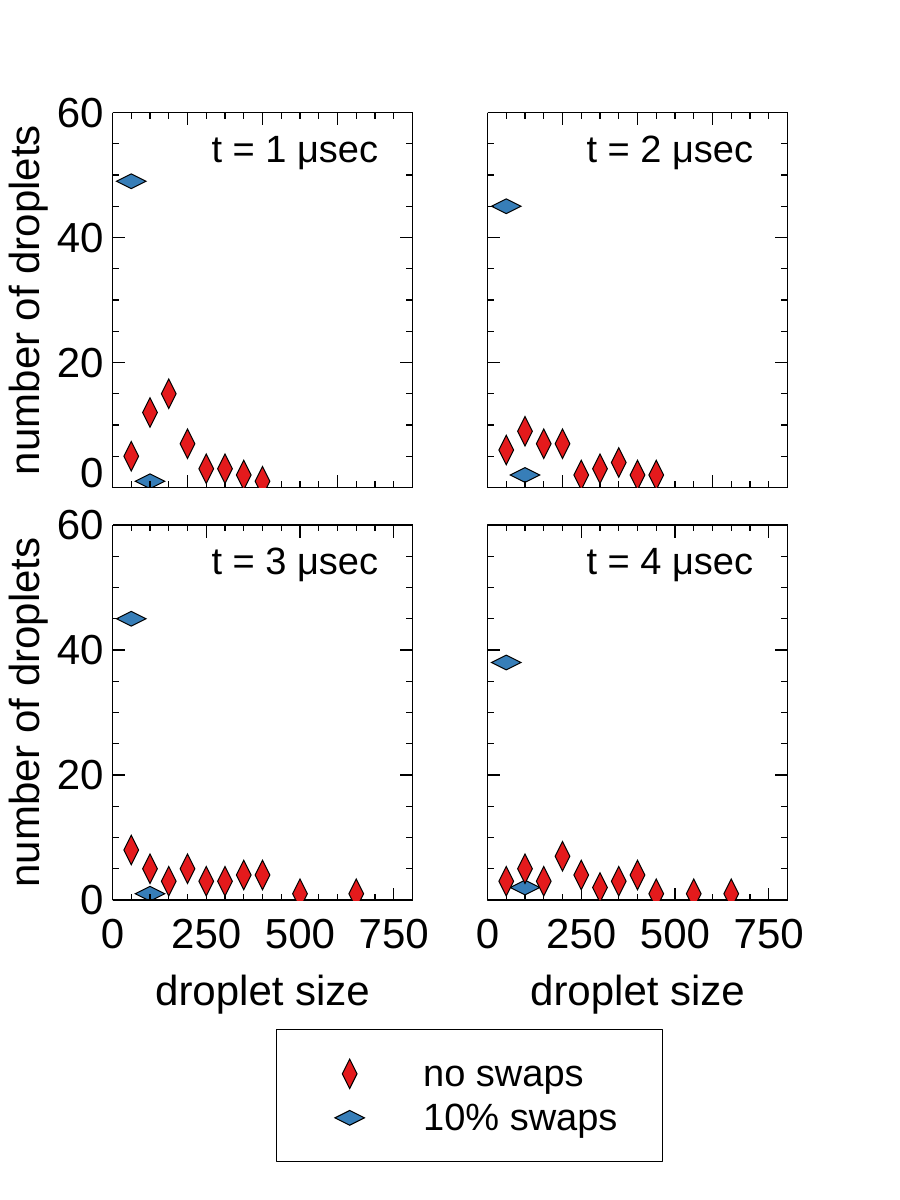}
\\[\smallskipamount] 
\caption{Same as Fig.~\protect\ref{fig:LJDW2p2},
with the double-well potential replaced by five times the Lennard-Jones potential.}
\label{fig:TwoLJMin}
\end{figure}

Simulations of the two-state models are shown in Fig.~\ref{fig:LJDW2p2} and Fig.~\ref{fig:TwoLJMin}. In each figure, the panels display the droplet size distribution at various stages of the simulation. In the equilibrium case, there is a clear trend toward increasing droplet size over time, consistent with Ostwald ripening. In contrast, in the non-equilibrium case, the system reaches a steady state characterized by a stable distribution of small droplets.

For all interaction potentials studied, we observe coarsening of the system in the absence of exchanges between \( A \) and \( B \) particles, and the formation of stable small droplets when the system is maintained out of equilibrium through \( A \leftrightarrow B \) exchanges.

For a system consisting of 70\% \( A \) and 30\% \( B \) particle states, the system evolves droplets that grow via coarsening and coalescence. When the transition rates \( k_1 \) and \( k_2 \) are small, we observe droplet formation followed by coarsening and coalescence. As the rates \( k_1 \) and \( k_2 \) increase, the state transitions begin to compete with droplet growth: particles in the \( B \) state are transformed into \( A \) and often leave the droplet, while particles in the \( A \) state are transformed into \( B \) and migrate toward the droplet. This dynamic exchange limits droplet size and stabilizes the system in a steady state with smaller droplets.

 In the limit of large transition rates, droplets cannot easily form, as \( B \) particle states in close proximity (i.e., a potential droplet) frequently transform into \( A \) particle states. The most common exchanges observed in the simulations involve the transition of a particle from a \( B \) state into an \( A \) state outside a droplet, and the transition of a particle from a \( B \) state into an \( A \) state inside a droplet.

The simulations show that in the limit of zero transition rates, the system behaves as a mixture of \( A \) and \( B \) particles undergoing coarsening and coalescence. In contrast, in the limit of very frequent transitions, no stable droplets are formed. The intermediate regime is studied here, and our simulations show that coarsening is suppressed (see Figures \ref{fig:LJDW2p2} and \ref{fig:TwoLJMin}) across a wide range of two-particle central interaction potentials. at the $10\%$ transition rate only small droplets are formed (blue markers in the figure). 

\begin{figure}[h]
\centering
\includegraphics[width=1.1\linewidth]{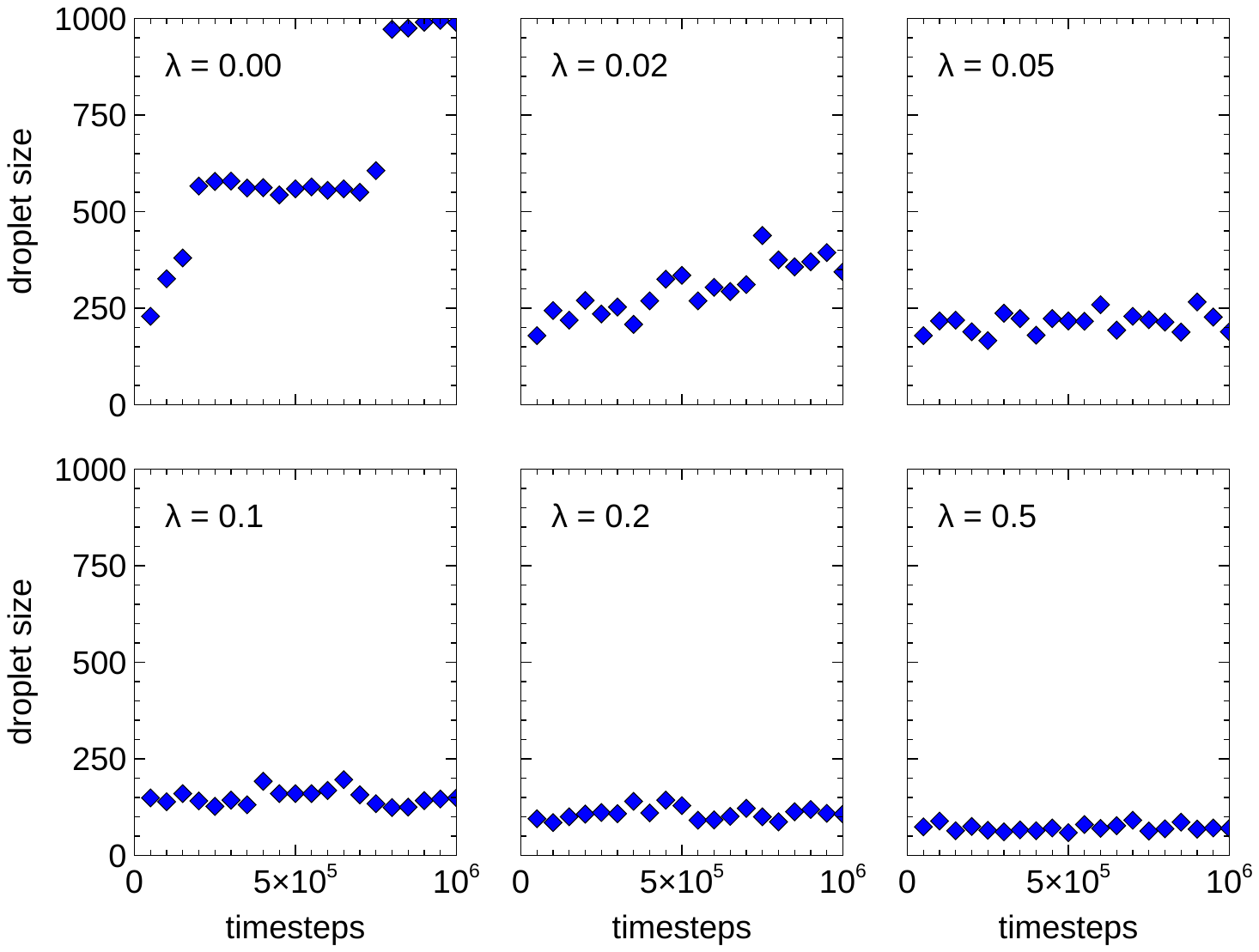}\hfill 
\caption{Plots showing the largest droplet size in the simulation as a
  function of time, for varying swap rate parameters $\lambda$ using
  the LJ+Gaussian model.}
\label{fig:lambda1}
\end{figure}

\begin{figure}[h]
\centering
\includegraphics[width=1.0\linewidth]{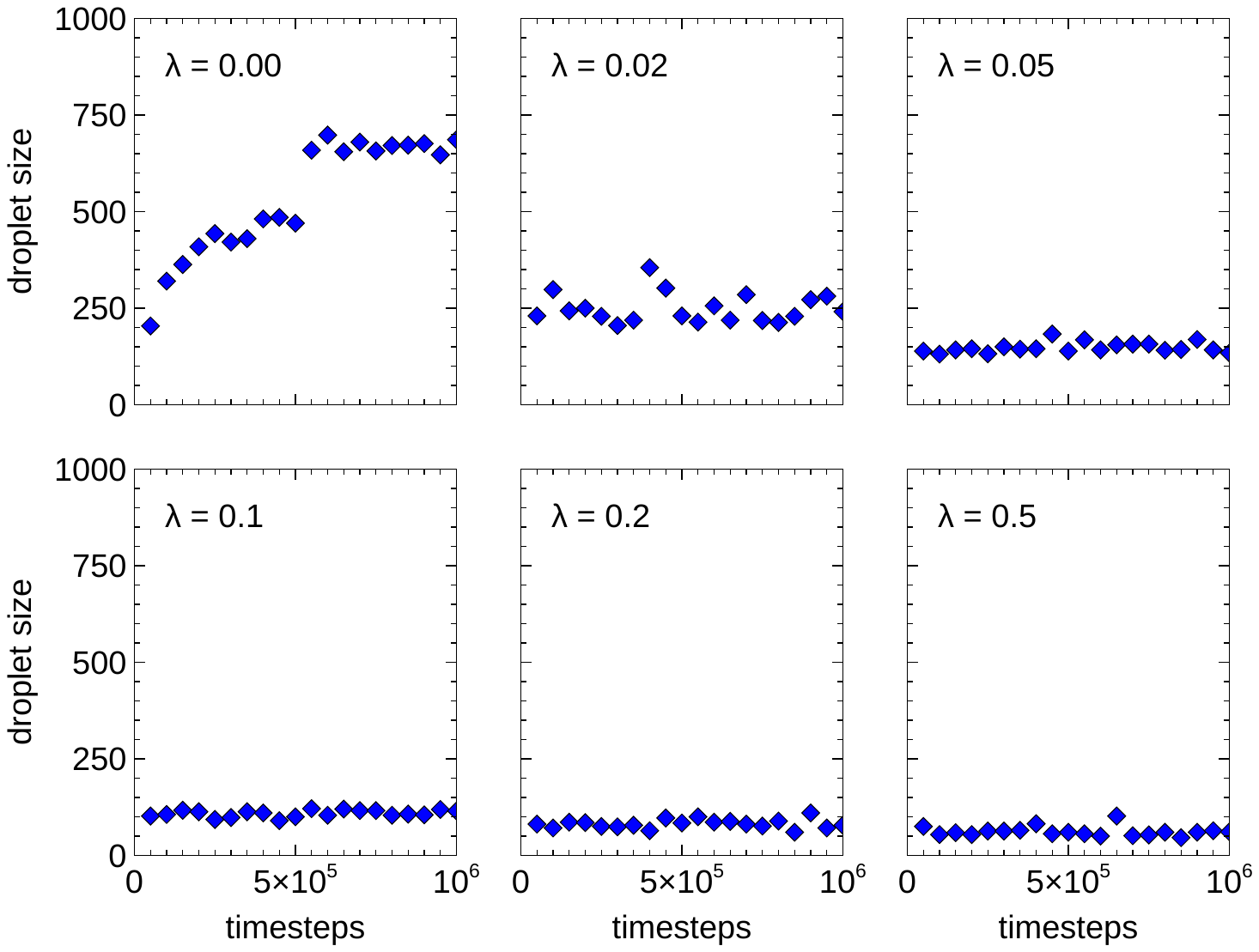}  
\caption{Plots showing the largest droplet size in the simulation as a
  function of time, for varying swap rate parameters $\lambda$ using
  the LJ times 5 model.}
\label{fig:lambda2}
\end{figure}

Figures~\ref{fig:lambda1} and~\ref{fig:lambda2} illustrate the extent of coarsening as a function of the swap rate, characterized by the parameter \( \lambda \), defined as the fraction of particles exchanged per 1,000 time steps. The plots show the size of the largest droplet as a function of time for various values of \( \lambda \).

When no swaps are performed (\( \lambda = 0 \)), the system continues to evolve, with droplets growing steadily (after one million time steps) likely due to coalescence. As the swap rate increasesthere
is very little droplet growth after a few hundred thousand steps, and
the droplet size becomes steadily smaller.


\begin{figure}[h]
  \centering
  \includegraphics[width=1.0\linewidth]{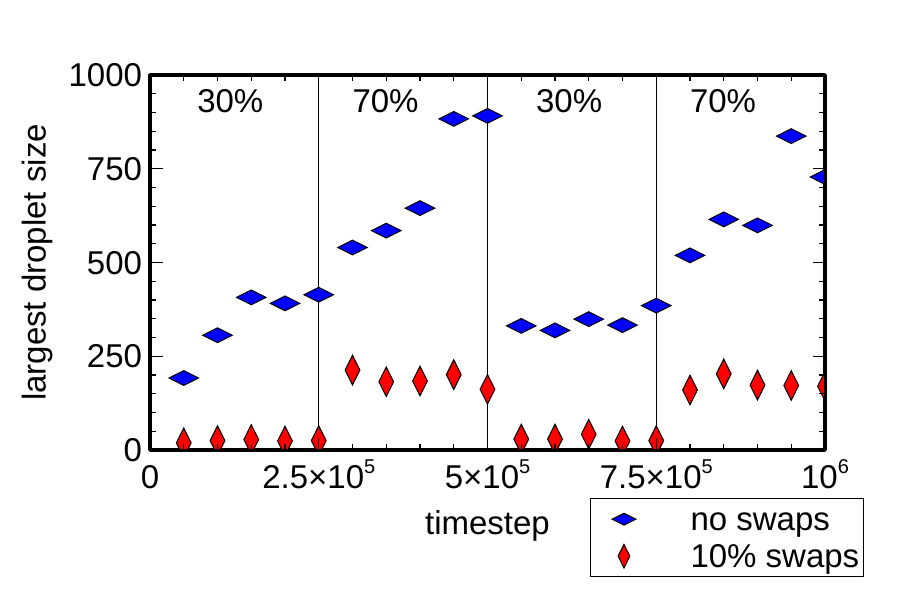}\hfill 
  \caption{Largest droplet size as a function of time, with the system
    alterrnating between 30\% and 70\% $B$ particles every 250000
    timesteps. The percentages shown in the figure correspond to the $B$ particles only. 
    The blue points correspond to no swaps, and the red points correspond to a 10\% swap rate.}
  \label{fig:alternate}
\end{figure}

We also examined the effect of varying the percentage of \( B \) particles in the system. Increasing the fraction leads to the formation of larger clusters but does not alter whether coarsening is arrested. Notably, if the fraction is changed during the course of a simulation, the system rapidly adjusts to the new configuration, as illustrated in Fig.~\ref{fig:alternate}.

In addition to driving the system out of equilibrium through transitions between molecular states—i.e., via changes in potential energy—we also studied a system driven out of equilibrium by kinetic energy. Specifically, we consider a system of identical single-state particles, with 30\% receiving random momentum "kicks" either parallel or anti-parallel to their instantaneous velocity. These perturbations are applied in such a way that the total energy of the system is conserved, but local equilibrium is disrupted.

\begin{figure}[h]
\centering
\includegraphics[width=1.1\linewidth]{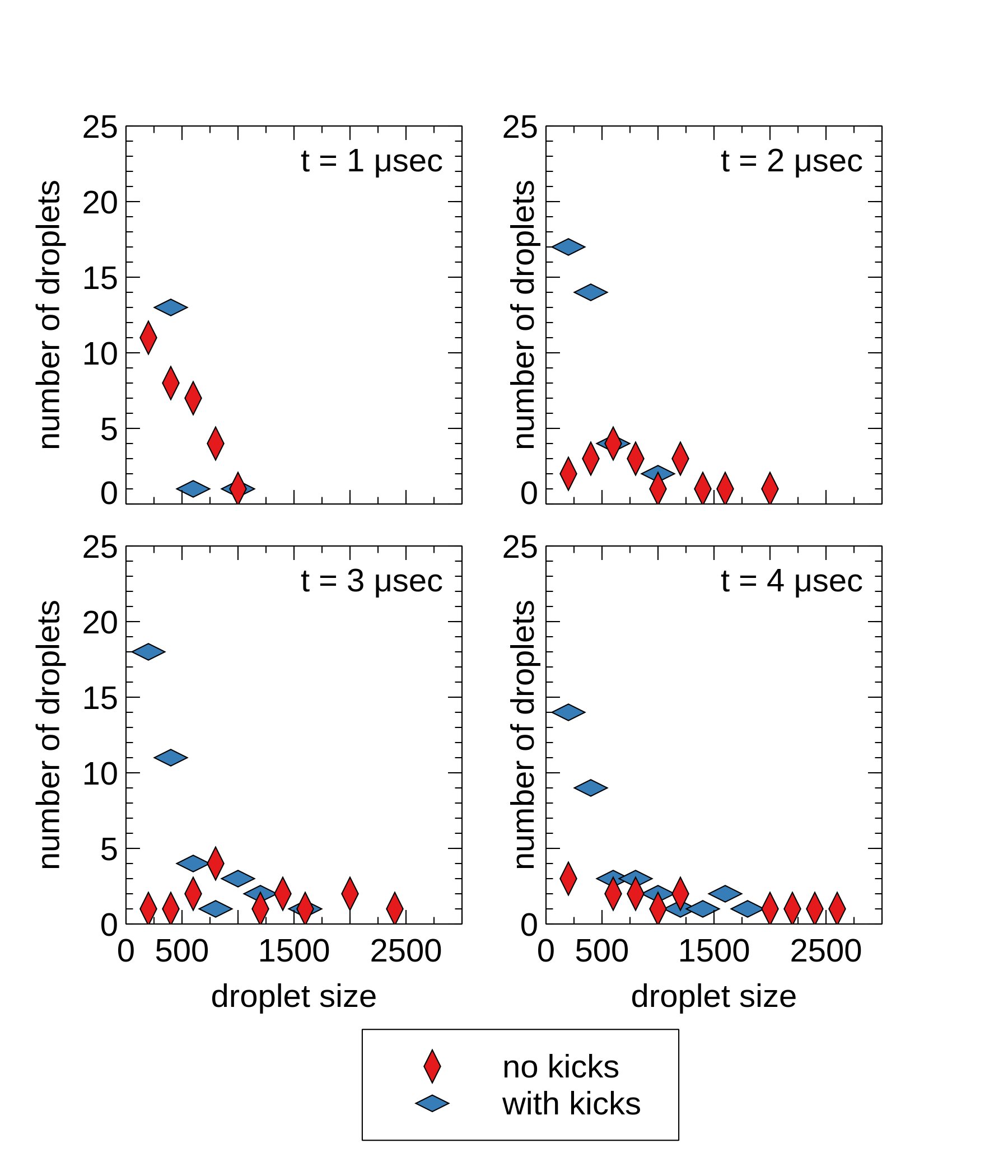}
\\[\smallskipamount] 
\caption{
  Same as Fig.~\protect\ref{fig:LJDW2p2}, with the
  non-equilibrium case represented by a simulation using random
  momentum 'kicks' to 30\% of the particles.}
\label{fig:Momentum}
\end{figure}

The results, shown in Fig.~\ref{fig:Momentum}, are qualitatively similar to those obtained in the potential-energy–driven system, with arrested coarsening and steady-state droplet formation.

\section{Analysis and Conclusions}

In this study, we used molecular dynamics simulations of many-body interacting systems with a selection of two-body potentials. Under equilibrium conditions, the system exhibits the expected coarsening behavior. However, when driven out of equilibrium, Ostwald ripening is suppressed and small droplets remain stable over long timescales.

A central feature of all simulations is the presence of local energy transfer—either via potential or kinetic energy—that maintains the system in a non-equilibrium steady state. In both cases, time-reversal symmetry is broken~\cite{Huberman1976}. Across all interaction potentials considered, we observed the absence of Ostwald ripening.

The timescale of our simulations is significantly longer than the timescale for initial droplet formation. Based on these observations, we expect that extending the simulations further would not qualitatively change the outcome: small droplets persist for long times, and eventual coarsening occurs through rare droplet coalescence. This coarsening process is expected to occur on a timescale that is long compared to physiological timescales, as it is limited by the slow diffusion of the stable droplets.

In equilibrium systems, Ostwald ripening arises because surface particles in small droplets experience a different net force compared to bulk particles. For small droplets, the surface-to-volume ratio is high, making surface effects dominant. As droplets grow, the surface-to-volume ratio decreases, stabilizing the larger droplets.

In contrast, when the system is driven locally out of equilibrium, energy deposition or extraction affects all particles. Since large droplets contain a greater number of bulk particles, they become energetically less favorable and tend to be destabilized. This effect is less pronounced for intermediate-sized droplets, which remain stable—consistent with our observations in this study.

Using local temperature sensors, it has been shown recently that different cellular compartments exhibit distinct temperatures~\cite{Pinol2020a}. These temperature differences are likely driven by ATP-supplied energy that is not fully dissipated, as well as by differing thermalization properties across organelles.

Our findings—showing the persistence of small droplets across a range of inter-particle potentials under non-equilibrium conditions—suggest that the non-equilibrium state of the cell may play a role in stabilizing small, membraneless organelles observed in many molecular assemblies. 

Beyond living cells, these results may also have implications for understanding other active matter systems driven far from equilibrium.








The authors would like to thank Jos\'{e} Onuchic, Peter Littlewood, and Herbert Levine for the useful comments and
discussions.  We thank Thomas Underwood for assistance using the
DL\textunderscore_MONTE Monte Carlo code.
Krastan B. Blagoev was supported by the National Science Foundation, while working at the Foundation. Any opinion, finding and conclusions or recommendations expressed in this material are those of the authors and do not necessarily reflect the views of the National Science Foundation
\bibliography{KBBreferences.bib}




\end{document}